\documentstyle[12pt,epsf,aasms4]{article}
\input psfig
 
\begin{document}

\newcommand{\nuf}{$\nu F_{\nu}$ }
\newcommand{\fun}{$\nu F_{\nu}$ }
 
\newcommand{\mesz}{M\'{e}sz\'{a}ros\/\ }
 
\newcommand{\mtaa}{ }
\newcommand{\etal}{{\it et al.}}
\def\mt4{ }

\newcommand{\gamb}{[\gamma_*^2 \, B_{ps}]}
 
\def\syn{synchrotron\/\ }
\newcommand{\cen}[1]{\centerline{#1}}
\def\loe{\lower 0.6ex\hbox{${}\stackrel{<}{\sim}{}$}}
\def\goe{\lower 0.6ex\hbox{${}\stackrel{>}{\sim}{}$}}
\newcommand{\ergs}{\rm \;  erg \, s^{-1}}
\newcommand{\ggg}{$\gamma$}
\newcommand{\eee}{$e^{\pm}$}
\newcommand{\be}{\begin{equation}}
\newcommand{\en}{\end{equation}}
\newcommand{\nn}{\noindent}
 
%
\def\jref#1 #2 #3 #4 {{\par\noindent \hangindent=3em \hangafter=1
      \advance \rightskip by 0em #1, {\it#2}, {\bf#3}, #4.\par}}
\def\rref#1{{\par\noindent \hangindent=3em \hangafter=1
      \advance \rightskip by 0em #1.\par}}

\noindent

\vskip .2in 
 
\title{X-Ray Emission of  Gamma-Ray Bursts}
 
\author{M. TAVANI}
 
\affil{Columbia Astrophysics Laboratory, Columbia University}
\affil{New York, NY  10027}

\baselineskip 20pt

\vskip 2.2in
\cen{Submitted to the {\it Astrophysical Journal}: August 30, 1996} 
\cen{Accepted: October 28, 1996}
\cen{To appear in the April 10, 1997 issue, Vol. 474.}

\begin{abstract}
X-ray emission  can provide a crucial diagnostic of gamma-ray bursts (GRBs).
We calculate the X-ray and gamma-ray spectra of 
impulsive acceleration episodes 
related to GRB pulses. We use the \syn shock model (SSM) as a basis
of our calculations. We show that the current data on 
soft-to-hard emission ratios of GRB pulse emission  are in agreement with 
the SSM.
In particular, GRB pulse emission detected by  GINGA 
is in agreement  with the SSM low-energy spectra.
We  deduce that GINGA detected the majority of  bright
GRBs detectable by BATSE.
These results indicate that the physical environment surrounding the 
GRB emission site is optically thin to X-ray photon energies.
We also  calculate emission ratios in the {\it Einstein}, ROSAT,
SAX and  HETE energy bands, and discuss
how future information on simultaneous soft/hard GRB emission 
can contribute in 
distinguishing  different emission models. 
Two different components of X-ray emission may simultaneously exist
in a fraction of GRBs.
One component is clearly associated with the individual GRB pulses,
and an additional component may be related to the pulse X-ray spectral upturns
and/or the precursors/tails occasionally observed.
We also show that a meaningful
 search of GRB-driven X-ray flashes in Andromeda (M31) can be carried out
with existing ROSAT PSPC data and future SAX WFC observations.

\end{abstract}

\vskip .2in
 

\newpage  %

\section{Introduction}

GRBs are characterized by a relatively low fluence in the
X-ray energy band as compared to the hard X-ray/gamma-ray
band (Trombka \etal\ 1974, Wheaton \etal\ 1973, Katoh \etal\
1984,
 Laros \etal\ 1984, Yoshida  \etal\ 1989, hereafter Y89,
Murakami \etal\ 1991). 
The soft/hard energy ratio (e.g., 1-10~keV/30-1000~keV)
of GRB fluences is of order of few percent.
 The ``X-ray paucity" feature is a
fundamental characteristic of GRBs, and a well established
observational fact.

We calculate in this paper the 
X-ray spectrum of GRBs
using the \syn shock model (SSM) which has been recently
shown to successfully reproduce the broad-band spectra of
bright GRBs (Tavani 1995;  Tavani 1996a,b,c,  hereafter T96a,b,c).
We show that in the absence of
absorption processes (due to opacity and/or \syn self-absorption)
in the X-ray band or spectral distorsions due to inverse Compton
scattering, the X-ray/gamma-ray emission ratios can be
reliably computed.
The observed emission ratios are dependent on the underlying 
particle energy distribution function,
and we calculate the X-ray/gamma-ray emission ratios
for different spectral assumptions.

We compare our results with the previously determined simultaneous
X-ray/gamma-ray emission ratios from the joint data of
the XMOS {\it P78-1} and the {\it Pioneer Venus Satellite} (PVO)
instruments
(Laros \etal\ 1984), 
the GINGA GRB monitor (Y89), and the WATCH instrument
(Castro-Tirado 1994).
We also calculate observable emission ratios appropriate to current
high-energy missions, BATSE, ROSAT,  SAX and HETE.

This paper is organized as follows. Sect.~2 provides a summary
of the GRB emission model used here, and a discussion of the
most relevant  theoretical points addressed in the analysis. 

A first goal of our paper is to use simultaneous soft/hard emission ratios
to constrain the GRB emission mechanism.
We show  in Sect.~3 that a definite correspondence between
X-ray/gamma-ray emission ratios and peak energy of the \fun
spectrum ($E_p$) can be established.
Simultaneous broad-band spectroscopy of GRBs detected by
different instruments can provide useful information to test
the SSM.

Our second goal is to discuss the possible existence of an
additional low-energy  component in the GRB spectrum (Sect.~4).
This extra component, most likely observable in the X-ray
energy range, might be easily detectable as a low-energy
excess during  the GRB pulse emission, or as a  component
 preceding or following GRB main pulses. 
Our analysis allows to easily identify spectral components
additional to the SSM pulse emission.

We  discuss in Sect.~5 attenuation and absorption processes
possibly affecting the detection of X-ray from GRBs.

We finally discuss in Sect.~6  the feasibility of a search for
GRB-driven X-ray flashes from other galaxies and in particular
from Andromeda. We consider ROSAT archival data  and
future SAX observations of Andromeda as examples of data
usable for this search.

\section{The SSM model}

The SSM is based on optically thin \syn emission of rapidly accelerated
relativistic particles
(electrons and/or \eee-pairs) radiating in the presence of a weak to moderate
magnetic field (to avoid magnetic absorption processes) (Tavani 1995,
T96a,b). A target `nebular' medium, able to reprocess the relativistic energy
of the flow and to trigger rapid acceleration processes, is necessary.
The ultimate origin of GRB magnetized relativistic particle flows 
can be  compact star coalescences at cosmological
distances or  compact star outbursts in an extended Galactic halo
(for a recent review, see Fishman 1996).
The target medium can be the interstellar medium, gaseous circumstellar material
or self-generated gaseous environments.
SSM can be applied in its generality to both the cosmological and Galactic
interpretations of GRBs, even though important differences between these
two models arise in  the radiation processes and overall dynamics 
(Tavani 1996d, hereafter T96d).
An MHD wind is assumed to interact in an optically-thin environment with magnetic
turbulence or  hydromagnetic shocks leading  to
rapid  particle acceleration and
to the formation of a prominent supra-thermal component.
The SSM relevant 
physical quantities are the  particle pre-acceleration `temperature'
 or  average  Lorentz factor $\gamma_*$, and  the local
magnetic field at the acceleration site $B_{s}$.
The  relativistic \syn
critical energy of emitted photons $E_c = h \, \nu_c^*$ 
(with $h$ Planck's constant and $\nu_c^*$ the critical
\syn frequency)
turns out to be proportional to the combination $[\gamma_*^2 \, B_s]$. 
A rapid
acceleration mechanism  of timescale
shorter than  the dynamic flow and cooling  timescales 
modifies an otherwise quasi-Maxwellian particle energy distribution (PED).
The post-acceleration PED, 
$N(\gamma)$ (with $\gamma$ the particles' Lorentz factor),
 turns out to be a combination of a
relativistic Maxwellian\footnote{
We assume a three-dimensional Maxwellian
distribution  valid for a randomly oriented magnetic
field configuration.
It can be shown that a two-dimensional distribution leads to 
results similar to those presented here for a randomly oriented
 magnetic field (T96b).}
and a power law component of index $\delta$  for energies below  and above
 $\sim \gamma_*$, respectively.
Depending on the efficiency of the acceleration mechanism,
the PED
 can have different shapes (T96a,b).
A {\it maximally efficient } acceleration mechanism 
is characterized by the non-thermal power-law component of the
 post-acceleration particle energy distribution joining
the
low-energy Maxwellian {\it at the top} of the distribution\footnote{
It can be shown that any other combinations of low-energy thermal and
supra-thermal components will lead to synchrotron/IC spectra
in contradiction with the current GRB broad-band data (T96a,b).}.
The SSM results in a dimensionless spectral function given by
 \be {\cal F} (w) \equiv \int_0^{y_c} y^2 \, e^{-y}
\; F'(w/y^2) \, d y  +
y_c^2 \, e^{-y_c} \,  
\int_{y_c}^{y_m} \, \left( \frac{y}{y_c} \right)^{-\delta}
\, F'(w/y^2) \, d y  \label{F} \en
where we defined
$w = \nu / (\nu_c^* \, \sin \alpha), 
   y = \gamma/\gamma_*,  
 y_m = \gamma_m/\gamma_* $, 
with 
$\nu_c^* = (3/4 \, \pi) (q \, B_s / m_e \, c) \, \gamma_*^2$
the critical frequency of particles of mass $m_e$ and charge
$q$ radiating in a local magnetic field $B_s$,
$\alpha$ the average pitch angle,
$y_c$ a critical value of the dimensionless energy variable
 $y$, $\gamma_m$ the upper cutoff of the
post-acceleration distribution function, and 
 \be F'(x) \equiv x \, \int_x^{\infty} K_{5/3} (x') \, d x'  \en
the familiar  \syn spectral
function with $K_{5/3}(x')$ the modified Bessel function of order
$5/3$.  
By integrating over the solid angle and emission volume,
and after dividing by the square of the distance,
we obtain the differential energy flux
$F^s_{\nu}$, 
i.e.,
\be F^s_{\nu}  \propto  {\cal F} (\nu/\nu_c^*) [\nu_c^*, \delta, y_c]  \en
where we made esplicit the dependence on the quantities
$\nu_c^*, \delta, y_c$
(we assumed the relation $y_m \gg y_c$, T96a,b). 
In the following, we use
$h \, \nu = E$ for the emitted photon energy.

Eq.~\ref{F} has a clear interpretation in terms of \syn radiation
of impulsively accelerated particles by a maximally efficient
mechanism. The  main property of particle acceleration is
reflected in the value of the critical dimensionless energy
variable $y_c$. If $y_c \sim 1$ ({\it model A)}, the acceleration process is
very rapid within the (comoving) dynamical timescale of the 
radiation front,  with a drastic depletion of the quasi-Maxwellian 
maximum of the pre-acceleration PED
near $\gamma_*$.  In this case, the acceleration
involves the majority of particles out of an initial quasi-thermal 
`reservoir'. 
On the other hand, an acceleration involving only a relatively small 
number of particles and producing a non-thermal tail of an
otherwise quasi-Maxwellian PED
 near $\gamma_*$ {\it (model B})
gives  $y_c \sim 4-7$.
These two possibilities are quite distinct, and we argue below that
future
X-ray/gamma-ray simultaneous observations of GRBs can 
be used to distinguish different PEDs.
If $y_c \goe 10$, the relevant  PED
turns out to
be of a quasi-Maxwellian form, with no appreciable non-thermal
component. 
Fig.~1 shows an example of SSM  calculated  photon spectra  $
F_{\nu}/\nu$, energy spectra $F_{\nu}$ and 
spectral power per energy decade \fun for the two models as a function of the 
dimensionless photon energy $E/E_c$.
We have assumed $\delta = 5$ and $y_c =1$ for  model $A$ that
reproduces the broad-band spectra of GRB~910814 and 920622 (T96a,b).
Model $B$ of Fig.~1 is given by $\delta = 5$ and $y_c =7$.
We note that the calculated spectra for the two models
are not self-similar. On the contrary, the
underlying quasi-thermal peak near $\gamma_*$ of model {\it B} leads 
to a  spectral curvature above $E_c$ considerably different than
for model {\it A}. The peak energies of the \fun spectra are also
considerably different, being $E/E_c \simeq 1.5$ for model $A$
and $E/E_c \simeq 25$ for model $B$.

 T96a,b used $y_c=1$, i.e., model {\it A} and showed that the
broad-band spectra of all bright GRBs detected
simultaneously by BATSE, COMPTEL and EGRET
 are in agreement with the SSM calculated
spectra.  A purely Maxwellian spectrum is in strong disagreement with
observations (T96a,b), and  a model with $y_c  \sim$~a~few
may be marginally consistent with the data.
It is important to point out, that GRO broad-band spectra of
GRBs can be determined in the energy range 30~keV-100~MeV.
It is  then clear that extending the simultaneous 
 spectrum to photon energies below 30~keV  provides a crucial 
test of the model.
From Fig.~1 is also clear that the low-energy range 
 can be strongly affected by the non trivial spectral
`curvature' determined by the underlying PED.
The simple extrapolation
$F_{\nu} \sim \nu^{1/3}$ may {\it not}  hold in the X-ray band, 
especially for $E_p \sim 100$~keV (see Fig.~1)
and/or hard-to-soft spectral evolution of the GRB pulse. 

Typical observed values of the peak photon energy  of the 
\fun spectrum $E_p$ are in the range $100~{\rm keV} \loe E_p \loe 1$~MeV
(Band \etal\ 1993, Ford \etal\ 1995).
The peak energy $E_p$ represents a clear feature of the 
broad spectrum, and Fig.~1 shows that its relation with
the relativistic \syn energy $E_c$  depends on the underlying
model for $N(\gamma)$.
There is also a non trivial dependence of $E_p$  on the 
index $\delta$ (T96b), and in the following we will take
into account both of these effects.
Even though the ultimate interpretation of $E_p$ in
terms of the physical quantity  $[\gamma_*^2 \, B_s]$ is
subject to a model-dependent factor, $E_p$ characterizes
the observed GRB spectra in a useful way.
In the following, we will show the calculated emission ratios
as a function of the observable $E_p$.

\section{GRB X-ray emission ratios}

We calculated GRB  X-ray emission ratios for a variety of
assumptions regarding the post-acceleration PED.
As a reference model of emission, we choose a SSM model
with $\delta=5$ and $y_c = 1$ (model $A$).
A different choice of PED parameters would result in softness
ratios differing at most by a factor of a few tens of a percent
 compared to those shown in all the figures except for Fig.~4.
Fig.~2 shows the results of a calculation of  softness energy flux
ratios (SRs) for energy bands appropriate to
 several X-ray instruments, i.e.,
GINGA
$SR_{GINGA} =$ f(1.5-10~keV)/f(1.5-375~keV) (Y89),
the XMOS NRL/Los Alamos experiment
$SR_{XMOS} =$ f(3-10~keV)/f(30-2000~keV) 
 (Laros \etal\ 1984), the 
WATCH instrument
$SR_{WATCH} =$ f(6-15~keV)/f(15-100~keV) (Castro-Tirado 1994),
{\it Einstein} 
$SR_{Einstein}  =$ f(0.15-3.5~keV)/f(50-300~keV),
ROSAT $SR_{ROSAT}  =$ f(0.1-2.4~keV)/f(50-300~keV),
the {\it Wide Field Cameras} (WFCs) of SAX
$SR_{WFC/BATSE} = $ f(2-30~keV)/f(50-300~keV),
 and the GRB {\it Wide Field X-ray Monitor} (WXM) of
HETE
$SR_{WXM/BATSE} = $ f(2-25~keV)/f(50-300~keV).
The quantity f  is the (arbitrarily normalized)
 integrated differential energy flux $F_{\nu}$
for the indicated  extremes of integration.
Note that the calculated emission ratios here and in the following
are idealized quantities that do not take into account possible spectral 
distortions due to detector responses.
We also neglect X-ray Galactic absorption  in our calculations of SRs.
X-ray absorption  has a negligible
effect for emission above 2~keV as detected by GRB monitors\footnote{
Obviously,  SRs for {\it Einstein} and ROSAT   may be strongly
affected by Galactic absorption.
Our calculations reported in Fig.~2 assume an unattenuated X-ray
flux. The reported SRs are therefore upper limits to the
true ratios.}.

The calculated SRs can be used to represent the softness
ratios  as a function of the {\it average} peak energy $E_p$ corresponding
to the GRB emission under investigation.
For example, the SRs can be representative of the emission near the
peak of a GRB pulse, or of the total fluence.
(In the latter case, $E_p$ represents
the average peak energy throughout the whole burst.)
Also the time evolution of the SRs
can be related with the change of $E_p$.

We can compare our results with the GRB detections in the X-ray band
of GINGA and XMOS\footnote{
WATCH softness ratios are currently not corrected for the aspect of the
detector (Castro-Tirado 1994), and a proper use of these data
requires more analysis.}.
The energy fluence ratios $SR_{XMOS}$ of bursts detected
between March and July 1979 are in the range 0.0096-0.034
(Laros \etal\ 1984). From Fig.~2 we deduce a range for the average
value of $E_p$ throughout the bursts, $200~{\rm keV} \loe E_p
\loe 500$~keV.
Yoshida \etal\ (1989) report values of $SR_{GINGA}$  referring to
the peak of the ten GRB bursts detected with good signal-to-noise
during  the period of March 1987 through March 1988.
They report values in the range $0.03 \loe SR_{GINGA} \loe 0.09$
(with the exception of one burst, 870319,  with $SR_{GINGA}=0.46$ that
would require an  `unusual' $E_p \simeq 10$~keV).
From Fig.~2 we therefore deduce a range of average $E_p$ for the bursts detected
by GINGA, $90~{\rm keV} \loe E_p \loe 300$~keV.

Three points are worth mentioning here.
The ranges of deduced $E_p$'s for both the XMOS and GINGA detections
are in good agreement with the expected average GRB $E_p$'s as determined
by BATSE (Band \etal\ 1993).
We deduce that the GRB emission of these bursts is in
agreement with the SSM expectations, with no necessity of additional
spectral components for the majority of bursts.
Only the burst 870319 detected by GINGA (Y89)
appears to have a SR
substantially larger than those expected from SSM spectra 
with $E_p \goe 100$~keV. 
A detailed spectral analysis can determine 
whether this burst has  either an anomalously low $E_p$,
 a prominent additional low-energy component, 
or the typical shape of  a soft gamma-ray
repeater event. A spectral analysis  of the 870319 burst is strongly
encouraged also in light of what discussed below.

The other important point to note here is that the
23 GRB detections  by GINGA for a total exposure
factor of $1.8 \, \rm sr \, yr$ (Teegarden 1995)
are consistent with the detection of 
the majority of {\it bright }
GRBs detectable by BATSE
(i.e., bursts with typical  peak fluxes above $F_t=(2.5-3) \rm \, ph \, cm^{-2}
\, s^{-1}$ averaged over 256~ms time bins, see Fishman \etal\ 1994,
Band \etal\ 1993). 
This can be derived from a comparison with
BATSE, that during the first 2.1~yr livetime period
detected 244 GRBs of spectroscopic quality
for an exposure factor of $\sim 18 \, \rm sr \, yr$
(e.g., Teegarden 1995). 
The ratio of the number of spectroscopic quality
 bursts to exposure is approximately the same
for GINGA and for the first 2.1 years of BATSE 
livetime.
This is an important result, indicating that
X-ray energy tails to GRBs are ubiquitous and with flux on the average
consistent with the SSM expectations.

The third important consequence  of these results is that
the environment surrounding the GRB emission site is
demonstrated to be
optically thin to  X-ray photon energies.
There is no evidence of absorption  processes in the X-ray
energy range for the majority of GRBs (see also Preece \etal\ 1996b).
There is also no evidence for 
  \syn self-absorption or substantial
inverse Compton distortions of the spectrum in the X-ray range.
These features are of great importance in  constraining theoretical models
(T96d).

Fig.~3 shows the calculated  hardness ratios (HRs) relevant 
for the BATSE energy channels 2 and 3, i.e.,
the energy flux $ HR_e =$ f(100-300~keV)/f(50-100~keV),
and the  photon flux $HR_p =  
{\rm f}_p({\rm 100-300~keV})/{\rm f}_p({\rm50-100~keV})$
with ${\rm f}_p$ the integrated photon flux $F_{\nu}/\nu$.
We notice that the majority of GRBs detected by BATSE have
average fluence HRs (e.g., Kouveliotou \etal\ 1993, 1996) 
in agreement with the SSM 
 calculated ratios.
Fig.~3 shows the dependence of emission ratios on the non-thermal high-energy
tail. We find that BATSE hardness ratios are not crucially dependent on the
PED.

Fig.~4 shows one of the main results of this paper, i.e., the 
SSM calculated ratio 
$SR_{SAX} =$f(2-30~keV)/f(60-600~keV) 
for a variety of PEDs.
Because of the relatively large energy span, $SR_{SAX} $ 
 depends on whether the PED is truncated near
$\gamma_*$ with $y_c \sim 1$ or quasi-thermal with
$y_c \goe 5$.
For the same $SR_{SAX} $, the peak energies $E_p$ deduced from
models $A$ and $B$ differ by a factor $\sim 2$.
Simultaneous broad-band spectral information in the
X-ray/soft \ggg-ray 
energy range can lead to a determination of the post-acceleration
PED. 
We note that our calculated softness ratios for the WFCs
can also be used to compare HETE's WXM data and BATSE data.
The SAX WFCs and HETE's WXM  may detect several GRBs per year,
and we can expect events detected simultaneously by any two of
 SAX, HETE and BATSE.
The broad-band spectral information of these bursts will be of
great importance.

We note that a further test of the SSM is provided by 
a comparison of GRB `pulse durations' $\tau_p$ for different energy bands.
A simple realization of the SSM   predicts  the 
relation $\tau_p \propto E^{-1/2}$ (T96a,b).   This relation is
in approximate agreement with auto-correlation
analyses of pulse durations by BATSE (Link \etal\ 1993, Fenimore \etal\ 1995).
Future data in the X-ray energy range can further constrain the
energy dependence of GRB pulse durations, and accurately measure
possible deviations from simple SSM predictions. 
Note that the GRB pulse duration can be quite different from the
duration of the extra X-ray component discussed below.

\section{Additional X-ray components}

We have shown that past and current X-ray
observations of GRBs are  on the average
in agreement with the SSM expectations.
However, a few exceptions were previously reported. 
Of particular relevance here, is the possible existence of 
{\it extra}  X-ray
components in addition to the SSM underlying spectrum.
One  extra component  occasionally  shows up
as an `X-ray excess'  during part of the bursts (e.g., Preece \etal\ 1996a,b).
The  existence of X-ray precursors and tails (with a possibly
different spectrum compared to the most intense part of the bursts) was 
 also reported
for the GRBs detected by GINGA
(Y89; Murakami \etal\ 1991).
It is not clear if the X-ray excess detected by BATSE
and the precursor/tail detected by GINGA have the same origin.
An analysis of the BATSE spectroscopy detector (SD) data  low-energy channel
is consistent with the presence of
 a statistically significant
excess in the 5-10~keV band in 12 out  of 86 strong bursts
 (Preece \etal\ 1996a,b).
The existence of GRBs with relatively large cumulative softness ratios  
was also noted in $\sim 10$\% of the bursts detected by WATCH
(Castro-Tirado 1994).
The limited spectral information of  the BATSE SD and of
WATCH and the statistics of the events detected by GINGA
does not as yet
 allow a precise determination of the spectrum of this
additional X-ray component. 
It is possible that GRBs imaged by HETE's WXM and 
 the WFCs on board of SAX
can further constrain this low-energy component.
GRBs with extra low-energy component(s)  would show softness ratios
substantially larger than those calculated for the SSM in
Figs.~2 and 4. 

A low-energy additional component of 
the GRB spectrum can have different physical origins
including {\it (i)} the existence of a quasi-thermal extra component of
temperature in the keV range, {\it (ii)} substantial spectral
hard-to-soft  evolution to peak energies $E_p \loe 10$~keV,
{\it (iii)}  the effect of opacity surrounding the central source
for Compton attenuation models (Brainerd 1994).
Only future detailed (time-resolved)  spectroscopy in the X-ray range
can resolve this issue.
We emphasize here that the presence of an extra component may
be crucial to distinguish different theoretical models.
Cosmological blast-wave scenarios may lead to observable 
quasi-thermal X-ray precursors and discrete X-ray emission  episodes
depending on geometry and interaction of forward and reverse shocks
(T96d). Surface re-emission of irradiated compact stars
in extended halo Galactic models
can also produce a quasi-thermal X-ray component which 
might  be delayed in time  with respect to the  main
GRB pulse emission.
A detailed discussion of models for these extra   X-ray components in GRBs 
will be presented elsewhere.

Current data indicate the existence of
two different X-ray spectra components, one associated with the
GRB  pulses (and successfully modelled by the SSM over a broad energy range),
and an extra component.
The latter component can occasionally modify only part of the
 GRB emission
(as in the case of the accumulated  spectra of 3B920517 near its  pulse peak
from 7.6 to 8.8~s after trigger, Preece \etal\ 1996a,b),
or manifest itself as a precursor/tail of relatively soft spectrum
(Y89).

\section{Attenuation and absorption of X-rays from GRBs}

Two effects may suppress the observable X-ray emission of GRBs, i.e.,
(1) attenuation of X-rays due to propagation from source to the
Earth, and (2)
opacity and \syn self-absorption effects at the source.

In previous sections, we assumed that the effect of X-ray propagation
in our Galaxy is negligible.
This is justified for column densities
below $5\cdot 10^{22} \rm
 \, cm^{-2}$ and photon energies well above 2 ~keV.
However, X-ray propagation through dense regions of the Galactic disk
may  substantially affect the low-energy spectrum of GRBs.
Fig.~5 shows a typical SSM spectrum for $E_p = 150$~keV and
$\delta=5, y_c = 1$ attenuated at low energies by photonionization
of neutral gas through column densities in the range $10^{21} - 5\cdot 10^{23}
\, \rm cm^{-2}$.

Detecting GRBs below 2~keV may result in valuable information
for distinguishing Galactic and extragalactic models of emission
(see also Schaefer 1994). 
The existence  of strong photoelectric absorption for GRBs occurring near
the Galactic plane  would be strongly  suggestive of a remote origin in
an extended halo or at cosmological distances. On the other hand,
unattenuated GRB spectra down to, say, 0.1~keV would be highly
problematic  for cosmological models.

Neutral gas may exist near the GRB source, and the propagation of
X-rays can be affected by the presence of unionized gas surrounding the
source of high-energy emission.
Time variable photoelectric absorption may then occur in GRB sources
of large initial column densities  of neutral gas. As  energy from 
the GRB  source is progressively absorbed by  cold surrounding 
material, the absorption cutoff will be  shifted to lower photon energies
in a distinctive way. Fig.~5 shows that the effective
column density  at the source must be larger than $10^{22} \, \rm cm^{-2}$ to
affect the GRB spectrum above 2~keV. From the absence of substantial
X-ray  absorption in the available GINGA and XMOS data (see Sect. 3),
we deduce that typical {\it average} values of $N_H$ are constrained below
$10^{22} \rm \, cm^{-2}$.
Time-resolved spectral X-ray information  will be valuable in
further constraining  the gaseous surrounding of  GRB sources.
Fig.~5 can be used to predict the time evolution of the low-energy
spectrum as the column density of photoionizing material 
evolves. 

Synchrotron self-absorption may also suppress X-ray emission of GRBs.
By equating the (comoving) spectral intensity $I_{\nu'}$ with
the Rayleigh-Jeans part of a blackbody spectrum, we obtain
$I_{\nu'} =  2 \, (\nu'/c)^2 \,  \gamma_*' \, m_e \, c^2$.
This relation is satisfied for  a critical frequency $\nu'_{abs}$
at which self-absorption sets in.
Note that $\nu'$  and $\gamma_*'$ refer to the (comoving) reference
system  where most of the GRB radiation is produced.
For a relativistically  moving radiation front, observed and
comoving quantities are related  by $\nu' = \nu/\Gamma,
\gamma_*' = \gamma_*/\Gamma$, with $\Gamma$ the bulk Lorentz
factor of the flow.
The specific radiation energy density $u_{\nu'}$ is then 
$u_{\nu'} \simeq 3.4\cdot 10^4 \, \nu_{abs,1}^2 \, \gamma_{*,5} \,
\Gamma^{-3} \, \rm erg \, cm^{-3} \, Hz^{-1} $,
where $\nu_{abs,1}$ is the observed absorption frequency in
units of 10~keV, and $\gamma_{*,5} = \gamma_*/10^5$.
The comoving average radiation energy density can be then expressed as
$\bar{u}' \simeq 8.4 \cdot 10^{22}  \, \nu_{abs,1}^3 \, \gamma_{*,5} \,
\Gamma^{-3} \, \rm erg \, cm^{-3} $.
A \syn self-absorption critical frequency $\nu_{abs}$
 in the X-ray range would
therefore indicate extreme conditions of radiation.
Relativistic beaming by a moving radiation front would somewhat 
alleviate the requirement on the radiation energy density.
The distinctive roll-over  and asymptotic  form of the energy spectrum
$F_{\nu} \sim \nu^{5/2}$ at frequencies below $\nu_{abs}$ is 
distinguishable from photoionization effects at X-ray energies.
We note that  $\nu_{abs}$ may evolve in time towards smaller
values as the GRB pulse progresses, and time-resolved X-ray
spectra might  be used to study the evolution of the burst
energy density at the radiating site.

\section{Search for X-ray  flashes in Andromeda and other galaxies}

X-ray flashes associated with GRBs from other galaxies might
be detectable by X-ray instruments.
If GRBs originate in a Galactic extended halo, it is plausible
to expect the detection of X-ray flashes related to GRBs in the halo of nearby
galaxies, since typically X-ray detectors are more sensitive
than those employed at higher energies
  (e.g., Hamilton, Gotthelf \& Helfand 1996).
Indeed, the detection or lack of detection of X-ray flashes in nearby galaxies
such as Andromeda may provide a crucial test  to establish the
nature of GRBs (see also, Li, Fenimore \& Liang 1996).

We show here that the calculated and previously observed 
X-ray flux of GRBs can be used to strengthen the conclusions
 from these searches.  As an example, we consider a search for 
GRB-driven X-ray flashes in Andromeda (M31) to be
carried out with ROSAT archival data. We assume  a fractional area of the
halo including the
Andromeda galaxy $\eta = \eta_{-1}  10^{-1}$ for a $1^{\circ}$x$1^{\circ}$
field of view, and a (cumulative) livetime $\tau_l = \tau_6 \, 10^6$ sec.
We can then estimate the ROSAT exposure of the Andromeda halo
 in the (unconventional)
units of `halo x time' as $3\cdot 10^{-3} \tau_6 \, \eta_{-1}
\, \rm halo \,  yr$.
The ratio of this exposure with
 the GINGA exposure
of the Galactic halo
 (re-expressed in the proper units as $\sim 0.14 \, \rm  halo \, yr $) is 
$\sim 2\cdot 10^{-2} \tau_6 \, \eta_{-1}$.
We deduce that the number of events detectable by the PSPC
 would be below unity
if the ROSAT detections were limited to the same fraction of the bright
GRB population accessible to GINGA.
However, this may not be the case.
From Fig.~1 we  deduce that the fractional flux expected in the ROSAT
band for a burst detected by BATSE in the 50-300~keV range 
(without an X-ray excess) is
$\xi_{ROSAT} =  10^{-2} \, \xi_{-2}$ 
for $E_p \sim 200$~keV (i.e., at the average value of
the $E_p$ distribution from BATSE data, Band \etal\ 1993).
 Typical peak luminosities in the BATSE energy band are
 $10^{-6}-10^{-7}  \, \rm erg \, cm^{-2} \, s^{-1}$.
Displacing this  population of bursts detected by BATSE by a factor in
distance of $\sim 4-5$ (representing the approximate distance of the Galactic
halo to Andromeda's), we can estimate the  peak luminosities in the ROSAT
band in the range $5\cdot 10^{-10}-5\cdot 10^{-11} 
\, \rm erg \, cm^{-2} \, s^{-1}$. For a typical absorption in
the halo of Andromeda ($N_H \simeq 10^{21} \, \rm cm^{-2}$, e.g., 
Trinchieri \etal\ 1988)
and a photon spectral index less than unity, we deduce PSPC countrates
in the range $22.5$-$2.2 \, \rm cts \, s^{-1}$.
We therefore conclude that,
depending on the burst duration and intensity, ROSAT  can  
detect GRBs in Andromeda  of intensity a factor of $\sim 10$  lower than
the bright bursts  detectable by BATSE and GINGA. 
Unless the  (logN-logS) intrinsic fluence distribution of GRBs from Andromeda
is drastically quenched for small fluences,
the number of GRB-driven X-ray flashes
 may be up to a  few (times $\eta_{-1} \, \tau_6 \, \xi_{-2}$)
 in the current ROSAT PSPC data.
The timing properties of these transient events can make
them distinguishable against the background.
By considering realistic values of $\eta_{-1} \, \tau_6 \, \xi_{-2}$ in
the range
 0.1--0.5 (taking into account limitations in the exposure
and moderate X-ray absorption), we estimate the number of X-ray flashes
expected in the ROSAT database of Andromeda to be of order unity.
We also notice that ROSAT and similar instruments such as {\it Einstein}
are not expected  in this model to detect GRB-related flashes 
of galaxies at distances larger than 2-3~Mpc$\, 
[\tau_l/( 10^5 \rm \, s)]^{1/2}$.

We can also consider a search for GRB-driven X-ray flashes in Andromeda
by the WFCs on board of SAX.
The typical WFC $5\sigma$ detection above a background of
$\sim 20 \rm \, cts \, s^{-1}$
(Piro \etal\ 1995) gives 
$\sim 40 \rm \, cts \, s^{-1}$.
This countrate, interpreted as a peak flux for an integration
time of order of a few seconds, corresponds to an energy
 flux of $\sim 0.2$~Crab,
i.e., $\sim  5 \cdot 10^{-9} \rm \, erg \, cm^{-2} \, s^{-1}$ in the
2--30~keV band.
Typical peak fluxes for bright GRBs detected by BATSE ($F \goe F_t$)
can be translated at the Andromeda halo distance, giving
energy fluxes $\goe (2.5-5)\cdot 10^{-8} \rm \, erg \, cm^{-2} \, s^{-1}$
in the 50-300~keV band.
From Fig.~2 we deduce that the calculated SSM fraction of the peak flux in 
the WFC vs. BATSE energy ranges is $\xi_{WFC} \sim 0.2-0.15$
for average peak energies in the range of 200-300~keV.
For bright bursts with values of $E_p$ 
at their peaks  in the range 200-300~keV (which are about half of the
total number of bright bursts detected by BATSE, see Ford \etal\ 1995),
we deduce an SSM flux estimate of $(0.5-1) \cdot 10^{-8}
\rm \, erg \, cm^{-2} \, s^{-1}$ in the 2-30~keV energy band.
These bursts might be detectable by the WFCs\footnote{
Note that bright bursts with relatively large values of $E_p$
may not be detectable by the WFCs at the Andromeda distance.
The softness ratio  $\xi_{WFC}$ is below 0.1 for $E_p \goe 500$~keV,
see Fig.~2.}.
We can estimate the necessary livetime/exposure\footnote{
The large field of view of the WFCs ($20^{\circ}$x$20^{\circ}$)
ensures that the Andromeda halo can be monitored by single pointings.
Therefore, in this case $\eta = 1$ and the exposure is equal to
the livetime.}
 for WFC observations of
Andromeda as a function of a required number of X-ray flashes,
$N_f$.
If the Andromeda halo has a population of bright GRBs similar to that one
accessible to GINGA in our Galaxy, 
 we deduce a  livetime $\tau_{WFC}  \sim 30-40$~days~$(N_f/10)$.
The hypothesis of an extended galactic  halo origin for the GRBs 
can therefore be tested  by long  WFC observations of Andromeda.
Other nearby spiral galaxies are outside the
distance range accessible to the study proposed here.

\vskip .3in
The author thanks Jules Halpern and David Helfand for discussions.
Research partially supported by NASA grant NAG~5-2729.

\newpage \vskip .1in \nn

\centerline{\bf References}
 
\rref{Band, D., \etal, 1993,  ApJ, 413, 281}   

\rref{Brainerd, J.J., 1994, ApJ, 428, 21}

\rref{Castro-Tirado, A., 1994, Ph.D. Thesis, Univ. of Copenaghen}

 \rref{Fenimore, E.E., IN'T Zand, J.J.M., Norris, J.P., Bonnell, J.T.
 \& Nemiroff, R.J., 1995, ApJ, 448, L101}

\rref{Fishman, G.J., \etal, 1994, ApJS, 92, 229}
 
 \rref{Fishman, G.J., 1996, in the Proceedings of the 3rd Huntsville Symposium
on Gamma-Ray Bursts, AIP, in press}

 \rref{Ford, L.A., \etal, 1995, ApJ, 439, 307}

\rref{Hamilton, T.T., Gotthelf, E.V. \& Helfand, D.J., 1996,
ApJ, 466, 795}

\rref{Katoh, M., \etal, 1984, in {\it High energy transients
in astrophysics}, ed. S.E. Woosley, AIP Conf. Proc. no. 115
(New York, American Institute of 
Physics), p. 390}
  
\rref{Kouveliotou, C.,  \etal, 1993, ApJ, 413, L101}

\rref{Kouveliotou, C.,  \etal, 1996,
Proceedings of the 3rd Huntsville Gamma-Ray Workshop, AIP,
in press}

 
\rref{Laros, J.G., Evans, W.D., Fenimore, E.E., Klebesadel, R.W.,
Shulman, S. \& Fritz, G., 1984, ApJ, 286, 681}

\rref{Li, H., Fenimore, E.E., Liang, E.P., 1996, ApJ, 461, L73}

\rref{Link, B., Epstein, R.I. \& Priedhorsky, W.C., 1993, ApJ,  408, L81}
 
\rref{Meegan, C., \etal, 1992, Nature,  355, 143}
 
\rref{Meegan, C., \etal, 1995, 3rd BATSE catalog, submitted to ApJ}
 
\rref{Murakami, T., \etal, 1991, Nature, 350, 592}
\rref{Preece, R., \etal, 1995,
{ Astrophys. \&  Space Science}, 231, 207}

\rref{Piro, L., \etal, 1995, SAX Observers' Handbook}

\rref{Preece, R., \etal, 1996a,
Proceedings of the 3rd Huntsville Gamma-Ray Workshop, AIP, 
in press} 

\rref{Preece, R., \etal, 1996b, ApJ, in press}

\rref{Schaefer, B.E., 1994, in Gamma-Ray Bursts,
Proc. 2nd Huntsville Workshop, eds. G.J. Fishman, J.J. Brainerd, K. Hurley
(New York, AIP Conf. Proceedings no. 307), p. 341}
 
\rref{Tavani, M., 1995, {Astrophys.  \& Space Science}, 231, 181}
 
\rref{Tavani, M., 1996a, Phys. Rev. Letters, 76, 3478 (T96a)}
 
 \rref{Tavani, M., 1996b, ApJ, 466, 768 (T96b)}
 
\rref{Tavani, M., 1996c, in the Proceedings of the 3rd Huntsville
Symposium on Gamma-Ray Bursts, AIP, in press}

\rref{Tavani, M., 1996d, in preparation}

\rref{Trinchieri, G., Fabbiano, G. \& Peres, G., 1988, ApJ, 325, 531}

\rref{Trombka, J.I. \etal, 1974, ApJ, 194, L27}

\rref{Wheaton, W.A., \etal, 1973, ApJ, 185, L57}

\rref{Yoshida, A., Murakami, T., Itoh, M., Nishimura, J.,
Tsuchiya, Fenimore, E.E., Klebesadel, R.W., Evans, W.D.,
Kondo, I. \& Kawai, N., 1989, PASJ, 41, 509}

\baselineskip 18pt

\newpage

\centerline{\bf Figure Captions}

\nn
FIG. 1 - 
Calculated (unattenuated) SSM spectra  for  models $A$ and $B$ with $\delta=5$.
{\it (Solid curves:)} model $A$ for $y_c = 1$;
{\it (dashed curves:)} model $B$ for $y_c = 7$.

\vskip .1in
\nn
FIG. 2 -
Calculated SSM softness ratios for energy ranges of 
different X-ray instruments.
Model $A$ with $\delta=5$ and $y_c=1$ is assumed with no
X-ray attenuation.
See text for the definition of energy bands.

\vskip .1in
\nn
FIG. 3 -
Calculated SSM energy and photon hardness ratios relevant for the
BATSE energy range.
{\it (Solid curves:)} Maxwellian PED;
{\it (short-dashed curves:)} non-thermal PED with $y_c=1$ and $\delta=5$;
{\it (long-dashed curves:)} non-thermal PED with $y_c=1$ and $\delta=4$;
{\it (dotted curves:)} non-thermal PED with $y_c=1$ and $\delta=3$.

\vskip .1in
\nn
FIG. 4 -
Calculated (unattenuated)
 SSM energy softness ratios relevant for the SAX WFCs and
lateral shields 
 for different assumed PEDs.
{\it (Solid curve:)} Maxwellian PED;
{\it (dotted-short-dashed curve:)} non-thermal PED with $y_c=7$ and $\delta=5$;
{\it (long-short-dashed curve:)} non-thermal PED with $y_c=5$ and $\delta=4$;
{\it (short-dashed curve:)} non-thermal PED with $y_c=1$ and $\delta=5$;
{\it (long-dashed curve:)} non-thermal PED with $y_c=1$ and $\delta=4$;
{\it (dotted curve:)} non-thermal PED with $y_c=1$ and $\delta=3$.
Calculated ratios for models $A$ and $B$ lead to clearly distinct
values of  the peak energy $E_p$.

\vskip .1in
\nn
FIG. 5 -
Photoabsorbed
 SSM photon spectrum  
in arbitrary units (model $ A$ with $y_c = 1, \delta = 5, E_p = 150$~keV)
for different equivalent column densities $N_H$ (solar abundance).
{\it (Dotted curve:)} unattenuated spectrum;
{\it (solid curve:)} photoabsorbed spectrum for $N_H = 10^{21} \rm \, cm^{-2}$;
{\it (long-short-dashed curve:)} $N_H = 5\cdot 10^{21} \rm \, cm^{-2}$;
{\it (dotted-long-dashed curve:)} $N_H = 10^{22} \rm \, cm^{-2}$;
{\it (dotted-short-dashed curve:)} $N_H = 5\cdot 10^{22} \rm \, cm^{-2}$;
{\it (long-dashed curve:)}  $N_H = 10^{23} \rm \, cm^{-2}$;
{\it (short-dashed curve:)} $N_H = 5\cdot 10^{23} \rm \, cm^{-2}$.

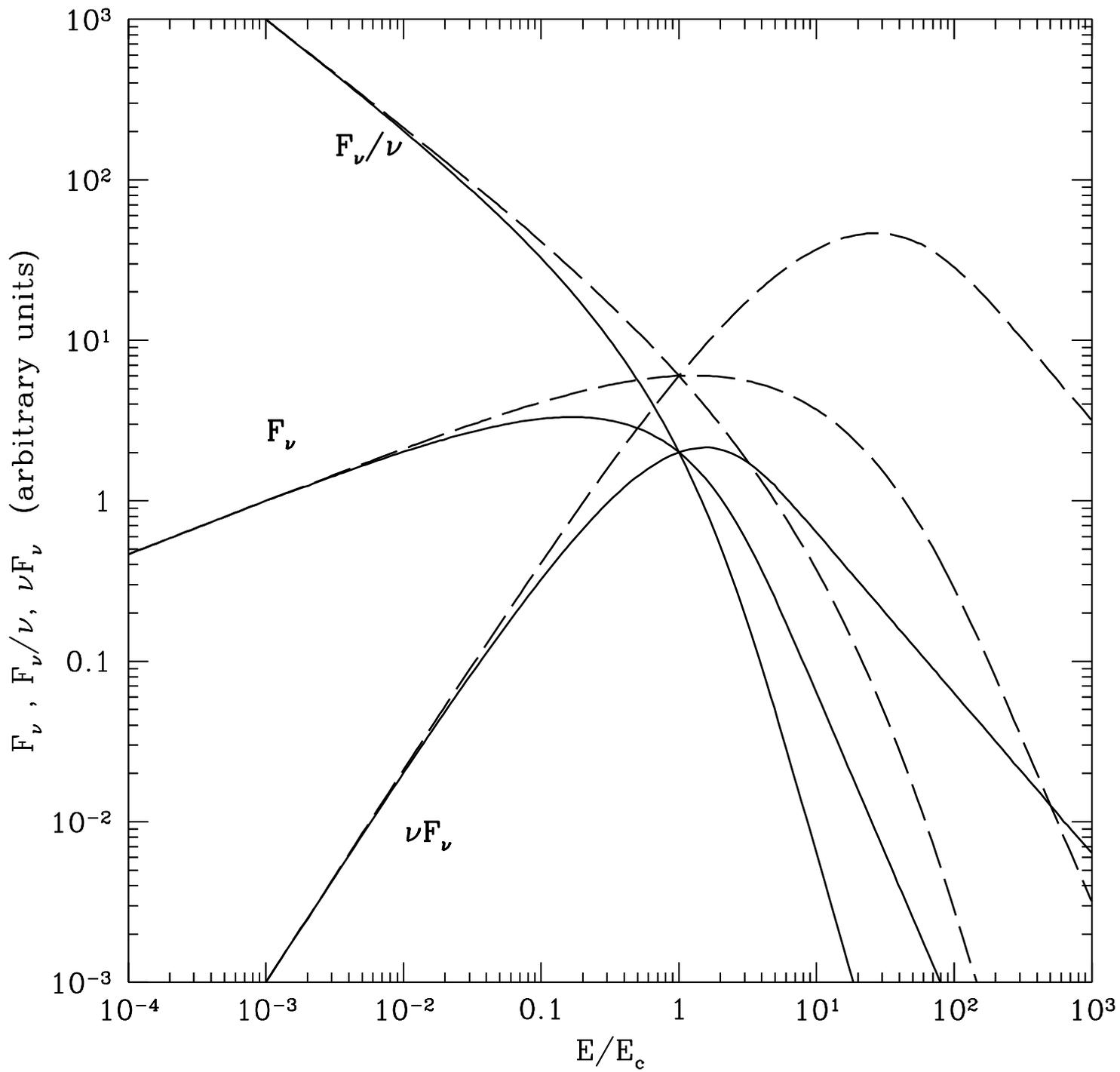
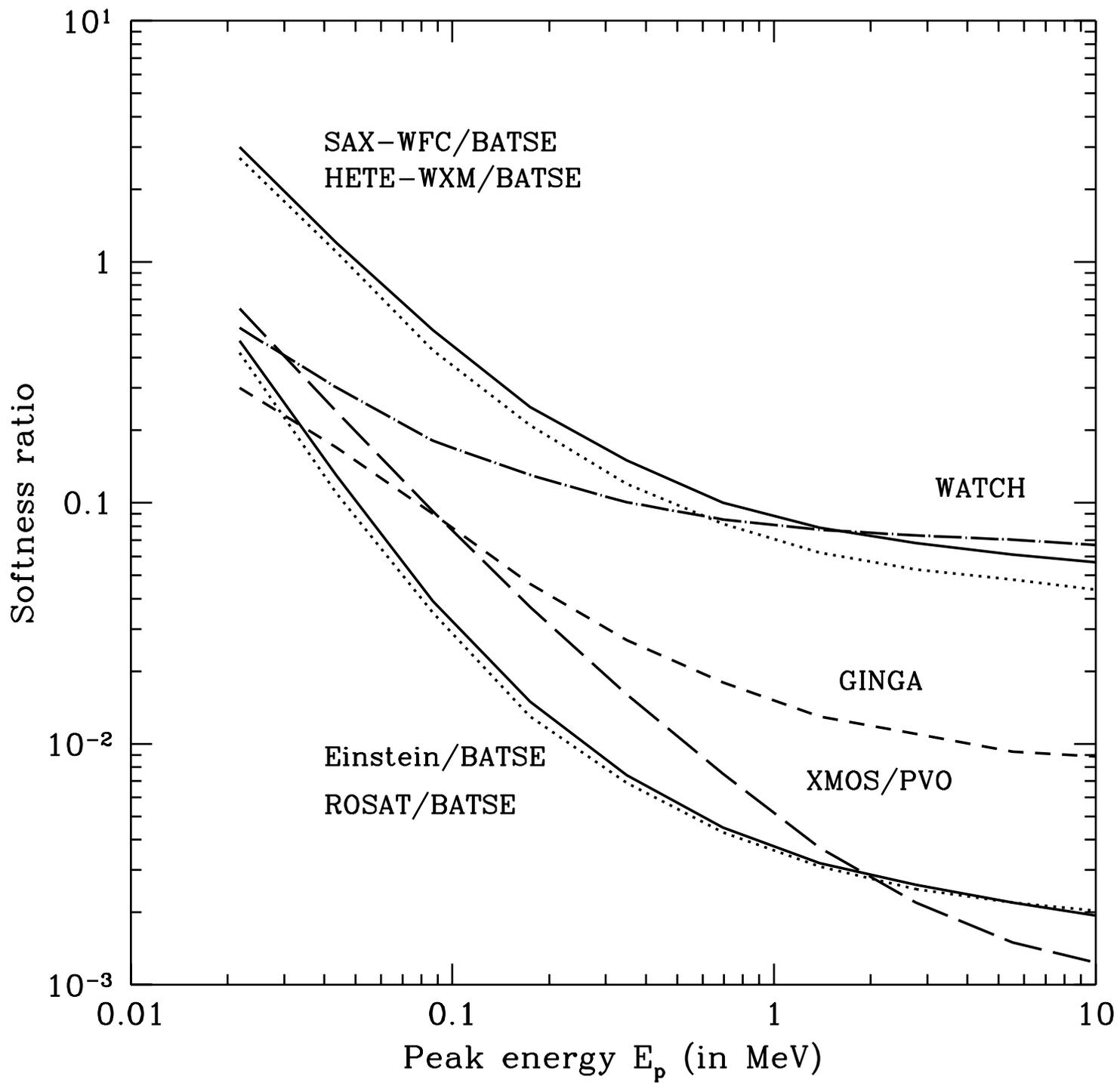
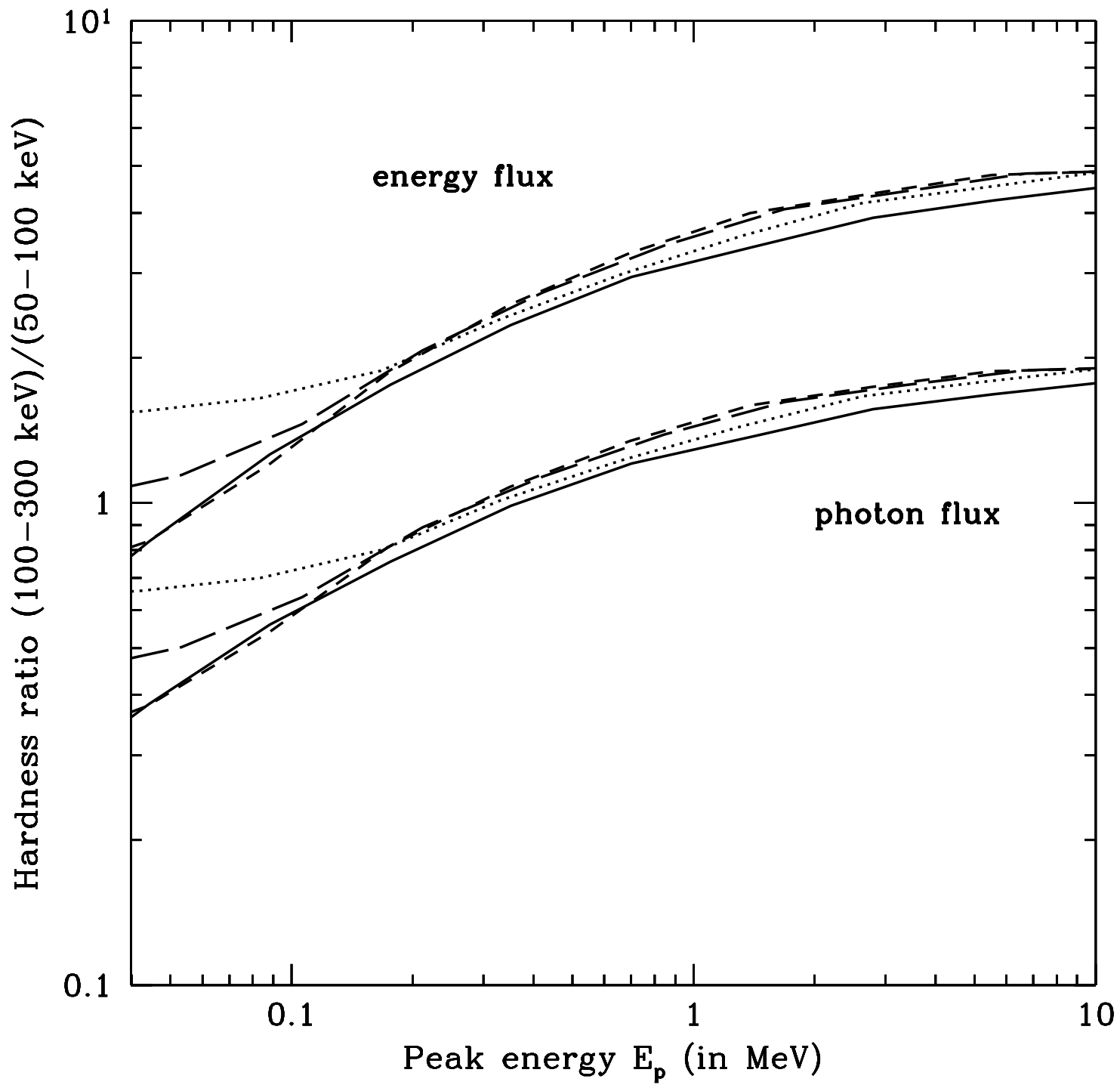
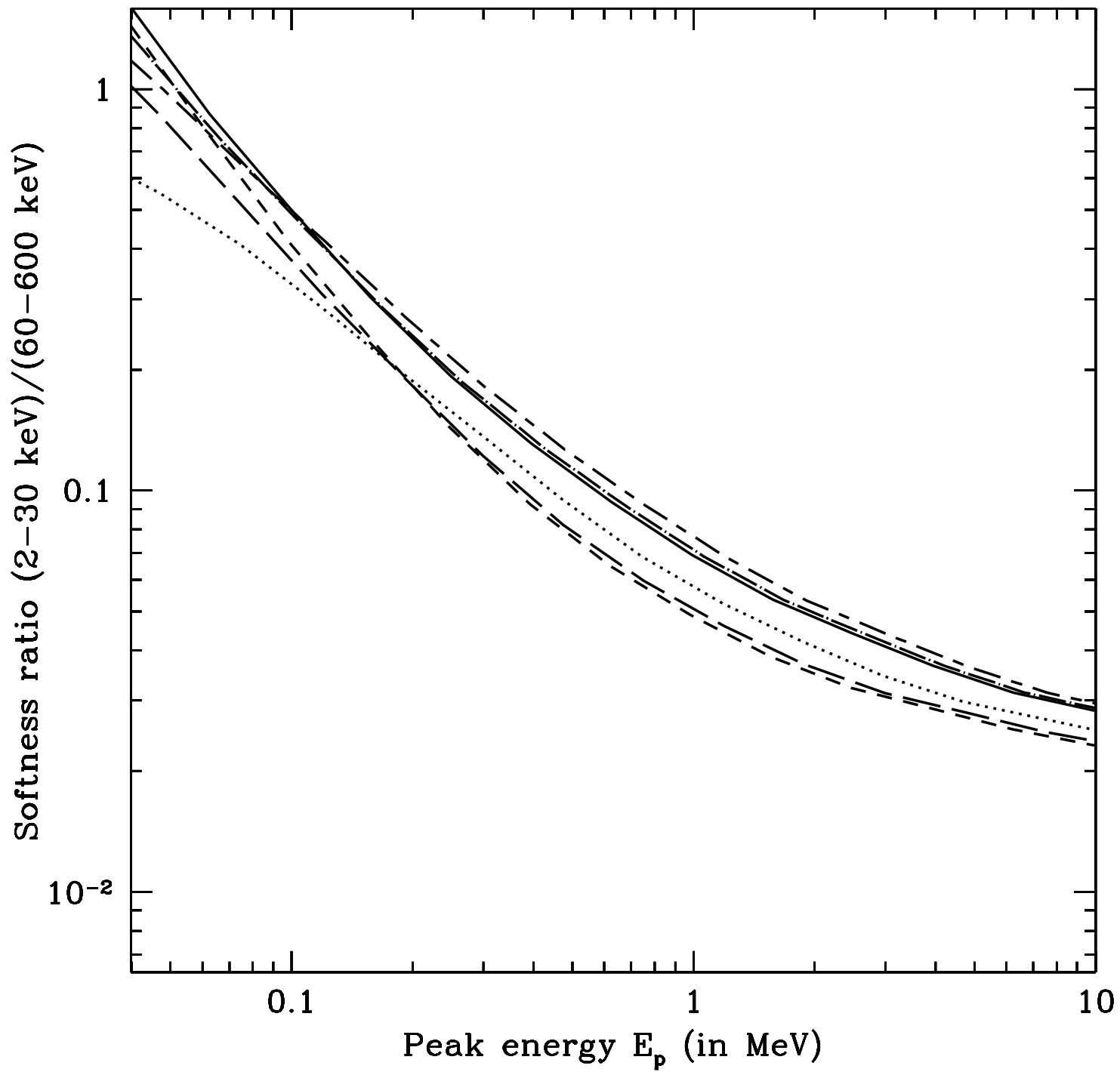
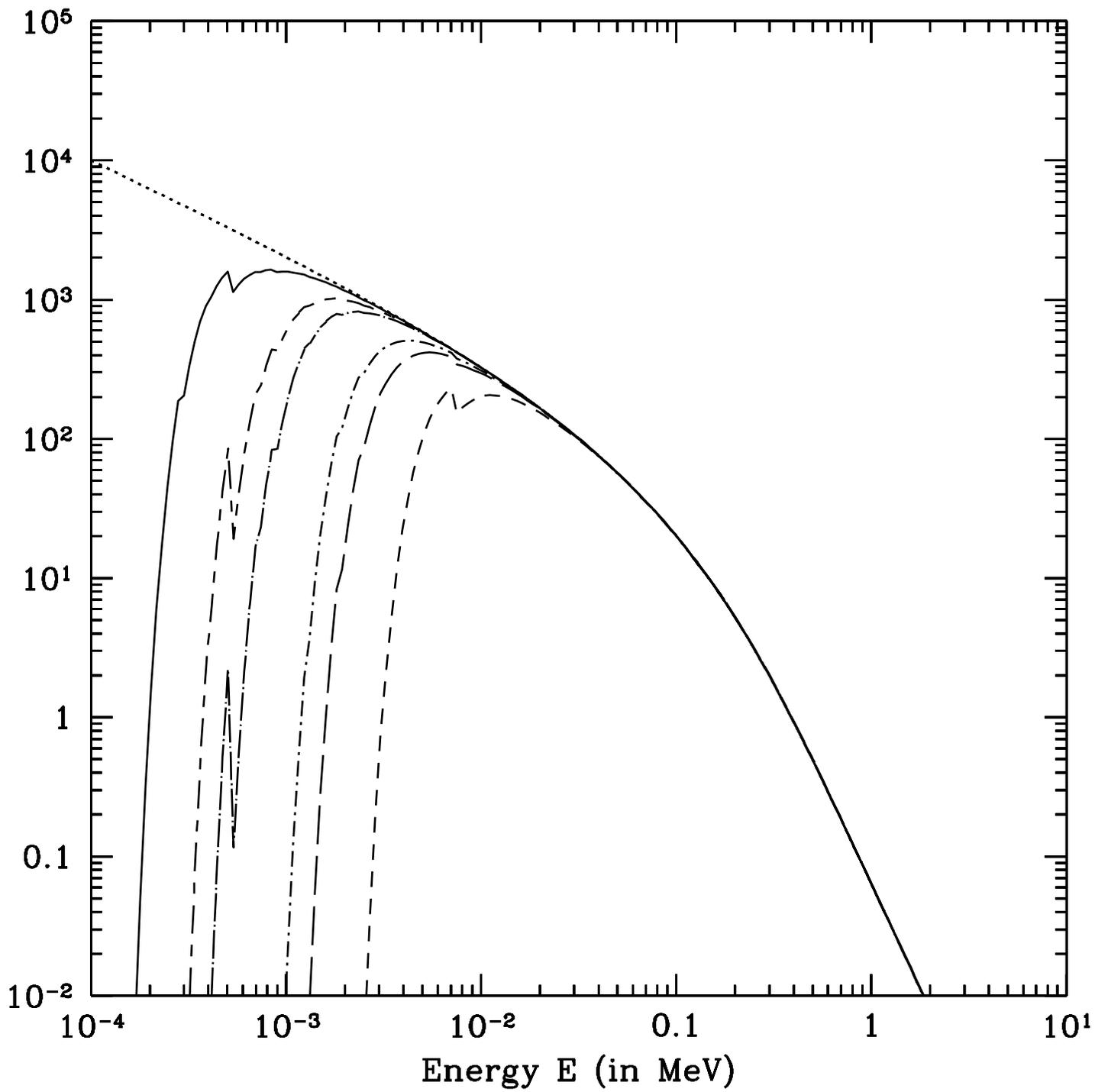
\end{document}